\documentclass[aps,prb,twocolumn]{revtex4}
\usepackage{graphicx}
\usepackage{amsmath}

	\newcommand{\ket}[1]{\left| #1 \right\rangle}
	\newcommand{\bra}[1]{\left\langle #1 \right|}

\begin{document}
\graphicspath{}

\title{Stability of topological properties of bismuth (111) bilayer}


\author{Maciej Bieniek}
\affiliation{Department of Theoretical Physics, Wroc\l aw University of Science and Technology, Wybrze\.ze Wyspia\'nskiego 27, 50-370 Wroc\l aw, Poland}
\author{Tomasz Wo\'zniak}
\affiliation{Department of Theoretical Physics, Wroc\l aw University of Science and Technology, Wybrze\.ze Wyspia\'nskiego 27, 50-370 Wroc\l aw, Poland}\author{Pawe\l ~Potasz}
\affiliation{Department of Theoretical Physics, Wroc\l aw University of Science and Technology, Wybrze\.ze Wyspia\'nskiego 27, 50-370 Wroc\l aw, Poland}


\date{\today}

\begin{abstract}
We investigate electronic and transport properties of bismuth (111) bilayer in the context of  stability of its topological properties against different perturbations. The effects of spin-orbit coupling variations, geometry relaxation and an interaction with a substrate are considered. Transport properties are studied in the presence of Anderson disorder. Band structure calculations are performed within multi-orbital tight-binding model and density functional theory methods. A band inversion process in bismuth (111) infinite bilayer and an evolution of edge states dispersion in ribbons as a function of spin-orbit coupling strength are analyzed. A significant change of orbital composition of the conduction and valence bands during a topological phase transition is observed. A topological phase is shown to be robust when the effect of geometry relaxation is taken into account. An interaction with a substrate has similar effect to an external perpendicular electric field. The robust quantized conductance is observed when the Fermi energy lies within the bulk energy gap, where only two counter-propagating edge states are present. For energies where the Fermi level crosses more in-gap states, a scattering is possible between channels lying close in $k-$space.  When the energy of edge states overlaps with bulk states, no topological protection is observed.
\end{abstract}


\maketitle

\section{Introduction}

Topological insulators (TI) are bulk insulators, which support protected boundary states \cite{4Moore, 5Hasan, 6Qi}. In two dimensional systems this is manifested by a presence of helical 1D edge states, with spin edge transport immune to non-magnetic disorder. This phenomenon is called quantum spin Hall effect (QSHE), and was first noticed theoretically for graphene with spin-orbit interaction included\cite{KanePRL}. It has stimulated large body of research on both 2D, and parallelly, 3D materials. There are several experimentally confirmed 2D topological insulators, including CdTe/HgTe/CdTe \cite{3Konig} and AlSb/InAs/GaSb/AlSb \cite{3Knez} quantum wells, ZrTe$_{5}$ \cite{3Li, 3Wu}, Bi~(110) \cite{3Lu} and Bi(111) bilayers \cite{1Yang, 1Sabater, 1Drozdov}, with many other systems predicted to support quantum spin Hall non-trivial state \cite{1Ren}. The list of experimentally addressed 3D topological insulators is even larger and constantly growing \cite{1Ando}.   

Topological protection against backscattering of electrons transported through edge channels was investigated by several authors\cite{ZhangPRB90, BaumPRL114}.  Situation is straightforward in case when pair of helical edge states linearly crosses the energy gap. Electrons propagating through edge channels are immune to backscattering due to spin-momentum locking. Since the time reversal symmetry is present in a system, it is not possible to flip the spin and change a moving direction. This leads to the conductance quantization in disordered samples up to the point when disorder couples edges of the system. On the other hand, an ideal situation with linearly dispersing edge states is not expected to occur in real systems. While topological insulators are characterized by an even number of edge states crossing the energy gap, it is not clear whether deviation from linear dispersion affects topological protection against backscattering. Although the effect of scattering between bulk and edge states was already studied for both 2D \cite{ZhangPRB90, BaumPRL114} and 3D TI's \cite{SahaPRB90}, the influence of extra in-gap states on transport is not well-understood. 

Bismuth (111) bilayer was one of the first real-life system predicted to be 2D TI \cite{MurakamiPRL97}. Free-standing bismuth bilayers have 0.2 eV band gap at the $\Gamma$ point and it was shown that it can be further increased to 0.8 eV by a proper choice of a substrate \cite{1Zhou}. This makes Bi~(111) very compelling as the gaps in other 2D TI's are of the order of meV. Calculations has shown QSHE in Bi(111) to be stable against strain and electric fields \cite{1Chen}, as well as choices of different substrates \cite{HuangPRB88}. Comparing to Kane-Mele \cite{KanePRL} and BHZ\cite{BernevigSCI} models, novel properties of edge states in ribbon geometry were recognized \cite{WadaPRB83}, and high level of tunability between localized and extended edge states by chemical means was predicted \cite{ChenRSC, JinSR, JinRSC, MaNano, MaPRB91, NiuPRB91, WangNano, WangRSC, LiPRB90}. Orbital magnetization in quantum dot geometry resulting from edge states circulation  was also shown to exhibit similar robustness \cite{PPNanoLett}.  

Despite intensive efforts to understand bismuth \cite{Hofmann}, some aspects of topological properties of this material are still not clearly understood. Although 3D Bi(111) crystal is conventionally known to be trivial insulator (opposite possibility discussed by \textcite{OhtsuboNJP}), non-trivial phase in thin films was shown to survive up to four bilayers without even-odd oscillations \cite{LiuPRL107}. The nature of edge states measured recently\cite{1Drozdov} in topmost layers of Bi (111)  is also controversial (\textcite{YeomPRB93} and references therein).  A problem of robustness of topological insulator phase in bismuth against different perturbations was also studied within a limited scope\cite{JinRSC, LiPRB90}.

In this work we focus on several aspects of Bi (111) bilayer as a realistic model of topological insulator. Using a combination of tight--binding and density functional theory methods, we study the effects of spin-orbit coupling parameter, geometry relaxation and an interaction with a substrate on the energy band structure. For 2D bismuth, a band inversion process as a function of spin-orbit coupling strength is examined. Ribbons are considered in two crystallographic orientations, with zigzag and armchair edges. We investigate changes in a dispersion of their edges states when perturbations are included. Next, the transport properties in pure and disordered systems are studied. By controlling the Fermi energy position, we analyze scattering processes  between the edge and the bulk states.  This allows us to distinguish energy regimes with and without topological protection against backscattering. We also verify whether the topological Anderson insulator phase \cite{11Li, 13Groth} exists for ribbons within a trivial regime, similarly to CdTe/HgTe/CdTe systems.

\section{Methodology}

\subsection{Lattice structure within Density Functional Theory method}
Bi(111) bilayer is a buckled 2D honeycomb crystal schematically shown in Figure \ref{fig:F1}(a). A hexagonal unit cell contains two atoms  and its geometrical parameters are:  a lattice constant $a$ and a bilayer thickness $h$. We conducted the 'ab--initio' calculations with ABINIT software \cite{Abinit}, which implements the density functional theory (DFT). The slabs of Bi bilayers and ribbons were separated by a vacuum region of $10$~\AA. The atoms were modeled within the frame of fully relativistic projector augmented waves (PAW)\cite{PAW} and the general gradient approximation (GGA) of the exchange-correlation functional \cite{GGA}. Structural parameters were optimized until the forces on atoms were smaller than $ 10^{-7}$ Ha/Bohr. The $8\times 8\times 1$ and $16\times 16\times 1$ Monkhorst--Pack k--point grids were used for structural optimization and density of states calculations, respectively. The plane wave basis cut-off was set to $20$ Ha.

Our DFT calculations give the following values of structural parameters: A-B atoms distance along z axis $h=1.73$~\AA\  and A-A atoms distance $a=4.43$~\AA. They are in a good agreement with calculations done by \textcite{HuangPRB88}, correspondingly $h=1.74$~\AA\ and $a=4.33$~\AA\ and experimental values from \textcite{HiraharaPRL109} ($h=1.64\pm 0.04$~\AA\ and a=4.39$\pm$0.05~\AA).  These values differ from parameters presented in \textcite{LiuPRL107} ($h=1.58$~\AA\ and $a=4.52$~\AA), although those calculations were performed with the use of the local density approximation, which is known to give different results for geometry optimization than GGA.

Besides the infinite Bi(111) bilayer plane, we have studied the electronic properties of ribbons, which are the finite width structures in a strip geometry periodic in one direction. We consider two most stable edge terminations of a honeycomb lattice, namely zigzag and armchair type of edges. The width of a ribbon  is determined by a number of atoms $N_{at}$ in a direction perpendicular to the edge. We investigate the bismuth (111) bilayer ribbons with $N_{at}=90$ atoms, unless otherwise stated.  

\subsection{Tight-binding method}
 We use four-orbital ($s,p_x,p_y,p_z$) tight-binding model (TB) developed by Liu and Allen \cite{LiuAllenPRB} for bulk bismuth with modification proposed by Murakami \cite{MurakamiPRL97}, in which all interlayer interactions are set to zero. The inter-atomic hopping up to the third nearest-neighbors and the atomic spin-orbit coupling (SOC) are parametrized with the Slater-Koster approach\cite{SlaterKoster}. Hamiltonian can be written as
 \begin{equation}
 \begin{aligned}
 &H=\sum_{i,\sigma,\vec{R}}\bigg( \ket{i,\sigma,\vec{R}} (E_{i}+U_{i})\bra{i,\sigma,\vec{R}} \bigg)+\\
 &\sum_{i,j,\sigma,\vec{R},\vec{R}^{\textrm{NN}}}\bigg( \ket{i,\sigma,\vec{R}} V_{ij}^{\textrm{NN}}\bra{j,\sigma,\vec{R}^{\textrm{NN}}}+\textrm{H.c.} \bigg)+\\
& \sum_{i,j,\sigma,\vec{R},\vec{R}^{\vec{NNN}}}\bigg( \ket{i,\sigma,\vec{R}} V_{ij}^{\textrm{NNN}} \bra{j,\sigma,\vec{R}^{\textrm{NNN}}} +\textrm{H.c.}\bigg)+\\
&+\sum_{i,\sigma,\sigma^{'},\vec{R}}\bigg( \ket{i,\sigma,\vec{R}} \frac{\lambda}{3} \vec{L}\cdot \vec{\sigma} \bra{i,\sigma',\vec{R}}+\textrm{H.c.} \bigg),
 \end{aligned}\label{Ham}
 \end{equation}
where $i,j$ indicate $\{s,p_x,p_y,p_z\}$ orbitals and $\{\sigma,\sigma'\}$ - spins. $\vec{R}$ are atomic positions with $NN$ and $NNN$ labeling nearest and next-nearest neighbors, $E_{i}$ denote on-site orbital energies and $U_i$ is on-site random potential. $V_{ij}$ are Slater-Koster two-center integrals between $i$ and $j$ orbitals. The last term corresponds to atomic SOC with $\lambda$ as SOC parameter. In our TB calculations, we first consider SOC strength as a fitting parameter and after comparison with DFT results presented in a next Section, we take $\lambda=1.8$ eV. This value is slightly modified in comparison with Liu and Allen $\lambda=1.5$ eV \cite{LiuAllenPRB}. Thus, TB parametrization may lead to minor variations of results, but they are all consistent in a qualitative way. 

In a first term of Hamiltonian given by Eq. (\ref{Ham}), $U_i$ is on-site random potential with values chosen from the uniform distribution from $\left[-\frac{W}{2},\frac{W}{2} \right]$. $W$ denotes disorder strength and is implemented in order to study transport properties in a presence of Anderson-type disorder. The conductance is studied as a function of disorder strength with results averaged over 100 disorder realizations, which, as we verified, was sufficient to obtain reasonably small statistical fluctuations.
 
\subsection{Transport calculations}
For transport calculations, we consider two-terminal geometry with semi-infinite leads attached to the left and the right edge of the scattering region. The Landauer formula for the differential conductance is given by
\begin{equation}
G=\frac{e^{2}}{h}\sum_{n\in L, m\in R}|S_{nm}|^{2},
\end{equation}
where $S_{nm}$ are the scattering matrix elements calculated using the recursive Green's functions method. $|S_{nm}|^{2}$ is calculated from
\begin{equation}
|S_{nm}|^{2} = \textrm{Tr}\left[\Gamma_{L}G^{r}_{1,N}\Gamma_{R}(G^{r}_{1,N})^{\dagger}\right]. 
\end{equation} 
where $G^{r}_{1,N}$ is a matrix representing the retarded Green's function between the first and the N-th slice, with slicing procedure presented in \textcite{Lewenkopf} for zigzag and armchair graphene ribbons. $\Gamma_{L(R)} $ is defined as a difference of semi-infinite lead self-energies $\left(\Gamma_{L(R)}~=\Sigma_{L(R)}~-~\Sigma_{L(R)}^{\dagger}\right)$, where electron self-energies are calculated using Sancho-Rubio\cite{SanchoRubio} iterative algorithm. All calculations were performed for non-interacting case and in $T=0$ K. Semi-infinite leads, attached to the edges of the system, were considered as made from the same material as studied system to avoid the contact resistance effect. We chose a system consisting of 90 by 180 bismuth atoms, which was sufficiently large to get size-independent results for the effects under consideration.

\section{Energy band structure}
\subsection{Bi(111) infinite bilayer}

\begin{figure}
\includegraphics[width=0.50\textwidth]{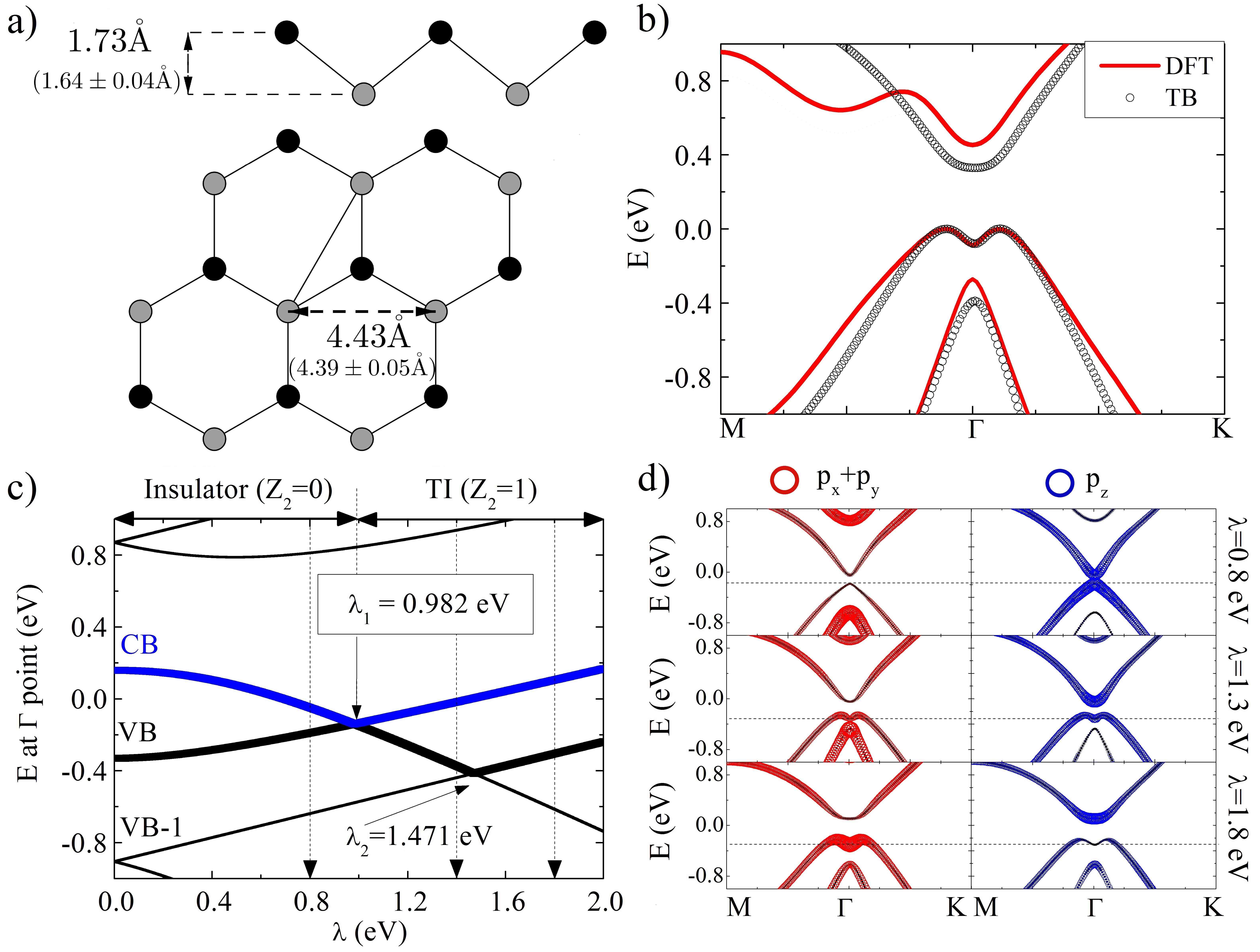}
\caption{\label{fig:F1}(a) Side and top view of bismuth (111) structure with parameters obtained from DFT calculations. Experimental values in brackets are taken from \textcite{HiraharaPRL109} and presented for comparison. (b) Energy band structures of Bi(111) bilayer along M-$\Gamma$-K direction obtained within DFT (red lines) and TB methods (black circles) with SOC $\lambda=1.8$ eV. (c) Evolution of energies at the $\Gamma$ point as a function of SOC parameter $\lambda$ using TB method. Topological phase transition is observed for $\lambda_1=0.982$ eV. The second band crossing between the valence bands is seen for $\lambda_2=1.471$ eV. (d) TB orbital composition for three different values of SOC parameter ($\lambda=0.8, 1.3, 1.8 $ eV). Radius of red (blue) dots represents $p_{x}+p_{y}$ ($p_{z}$) orbitals contribution to bands.}
\end{figure}

Low energy band structures of Bi(111) bilayer along M-$\Gamma$-K direction obtained within DFT and TB methods are shown in Fig. \ref{fig:F1}(b). We observe a  satisfying agreement between TB model and DFT results close to maximum of the valence band for SOC parameter $\lambda=1.8$ eV. The band structure has well defined energy gap at the $\Gamma$ point. Within TB method $E_{\textrm{gap}}\approx0.2$ eV, and even larger gap ($E_{\textrm{gap}}\approx0.4$ eV) from DFT calculations is obtained. We have verified non-trivial topology of band structure  by calculating  $Z_2$ invariant for inversion symmetric systems according to method from \textcite{FuPRB}. 

A band inversion point is determined by looking at evolution of valence band maximum and conduction band minimum as a function of SOC parameter within TB method \cite{acta2016}, shown in Fig. \ref{fig:F1}(c). Topological phase transition is observed for $\lambda_1=0.982$ eV. A further increase of SOC parameter leads to the second crossing within the valence bands at  $\lambda_2=1.471$ eV, but it does not change the $Z_2$ invariant. A splitting between two valence bands, labeled as $VB$ and $VB-1$, depends crucially on the SOC strength (Fig. \ref{fig:F1}(c)). A choice of $\lambda=1.8$ eV in our TB model with the band structure shown in Fig. \ref{fig:F1}(b) was motivated by fitting $VB$ to $VB-1$ splitting to our DFT calculations.

TB orbital composition of low energy band structure in a trivial phase for $\lambda=0.8$ eV, topological insulator phase before the second band crossing for $\lambda=1.3$ eV, and topological insulator phase after the second band crossing for $\lambda=1.8 $ eV, are shown in Fig. \ref{fig:F1}(d). For values of $\lambda$ corresponding to a trivial phase low energy bands are composed mostly of $p_z$ orbitals. With an increase of SOC strength, the top of the valence band looses  $p_z$ character and becomes composed also from $p_x+p_y$ orbitals. One can also note that with a variation of SOC parameter, a character of band gap changes, from direct (in a trivial phase) to slightly indirect (Fig. \ref{fig:F1}(b)) after topological phase transition. After the second band crossing for $\lambda_2=1.471$ eV, a $p_z$ orbital contribution to the top of the valence band is further decreased. 

\subsection{Bi(111) bilayer ribbons}
A characteristic feature of the electronic band structure of topological insulators in a strip geometry is the presence of counter-propagating edge states crossing the energy gap. We investigate stability of these edge states against a variety of perturbations.

\subsubsection{Spin-orbit coupling effect}
\begin{figure}
\includegraphics[width=0.50\textwidth]{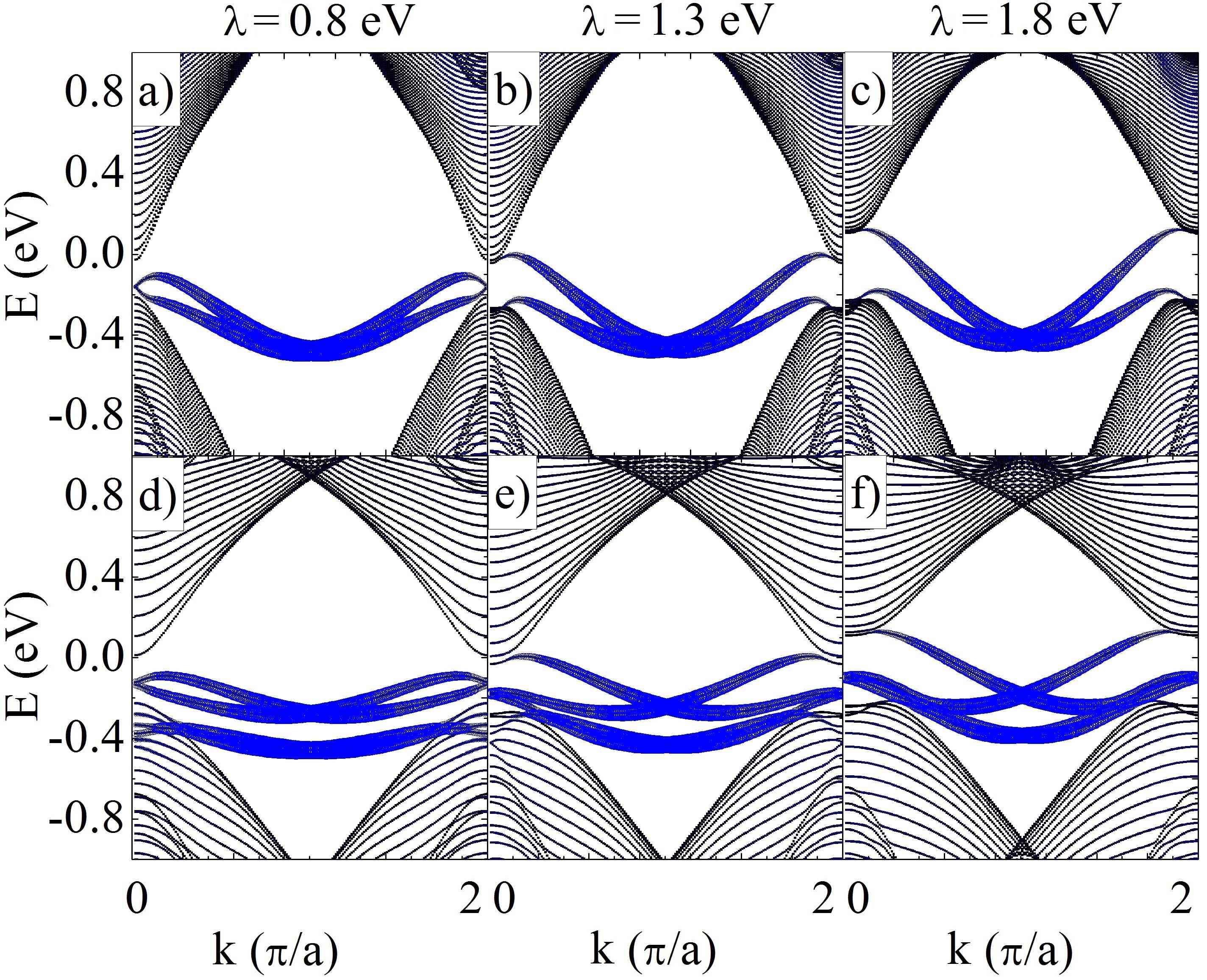}
\caption{\label{fig:F2} Energy band structures of zigzag (a -- c) and armchair (d -- f) bismuth ribbons with $N_{at}=90$ for three different values of the SOC parameter, $\lambda=0.8$ eV (a) and (d), $\lambda=1.3$ eV (b) and (e), and $\lambda = 1.8 $ eV (c) and (f), respectively. A thickness of the blue line represents a magnitude of the localization of states at two edge atomic sites. For $\lambda = 1.3$ eV and $\lambda = 1.8$ eV in both zigzag and armchair ribbons there are edge states connecting bulk valence and conduction bands, which is characteristic for topologically non-trivial states.}
\end{figure}

While SOC is characterized by a type of material, it can be partially controlled by external factors like dopants or lattice curvature \cite{Fabian, Fratini, Hernando}. We would like to analyze how SOC strength affects a dispersion of edge states. We consider wide ribbons to make the 1D band structure width independent. This allows us to focus on more general properties of the studied structures. In Fig. \ref{fig:F2}, energy band structures for ribbons with $N_{at}=90$ with zigzag (a-c) and armchair (d-f) edges are shown. We have considered three values of SOC parameter (as in Fig. \ref{fig:F1}(d)), one representing a trivial phase for $\lambda=0.8$ eV (a) and (d), topological insulator phase before the second band crossing for $\lambda=1.3$ eV (b) and (e), and topological insulator phase after the second band crossing for $\lambda=1.8 $ eV (c) and (f). A size of the blue lines in Fig. \ref{fig:F2} represents a magnitude of the localization of states at two atomic sites on both edges, showing that in-gap states are indeed localized almost solely on the boundaries of the ribbons.

First, we consider zigzag edge termination. For weak SOC $\lambda$=0.8 eV within a trivial phase, two branches of edge states are attached to the top of the valence band at the boundaries of the Brillouin zone. Their dispersion is quasi-flat and they do not cross the energy gap (Fig. \ref{fig:F2}(a)). An increase of SOC leads to a shift of an upper branch to the conduction band. After reaching a critical value of SOC, an upper branch of edge states touches the conduction band at the boundaries of the Brillouin zone. We relate it to a topological phase transition after which the material becomes TI. A further change of SOC increases an energy band gap which causes a larger dispersion of edge states, see Fig. \ref{fig:F2}(c). In a case of armchair ribbons, two branches of edge states are always present. For a weak SOC one pair of edge states lies within the energy gap and the second pair has lower  energy, overlapping with the valence band. Increasing SOC affects stronger the in-gap edge states, making them more dispersive. When a critical value of SOC is exceeded, they cross the energy gap connecting the valence and the conduction bands similarly to the zigzag ribbon case. The lower branch of edge states remains attached to the top of the valence band and is not strongly affected by the change of SOC. Thus, we can assume that only one branch of edge states has a topological origin, and the second one can be associated with trivial edge dangling bonds.

An interesting feature of topological edge states in bismuth, in both types of ribbon edge termination, is that their dispersions depart from linearity, contrary to clear Dirac-like spectrum in e.g. Kane-Mele model \cite{KanePRL}. Close to the top of the valence band, one can notice a double crossing of the given Fermi energy by edge states branch. A velocity direction of mobile electrons is determined by a dispersion relation $v\sim\delta E/\delta k$. Here, this quantity for edge states changes sign within the Brillouin zone. This may affect the conductance due to possible scattering between counter-propagating states within a given edge. We come back to this issue in Section IV.

\subsubsection{Edge relaxation and substrate effect}
\begin{figure}
\includegraphics[width=0.50\textwidth]{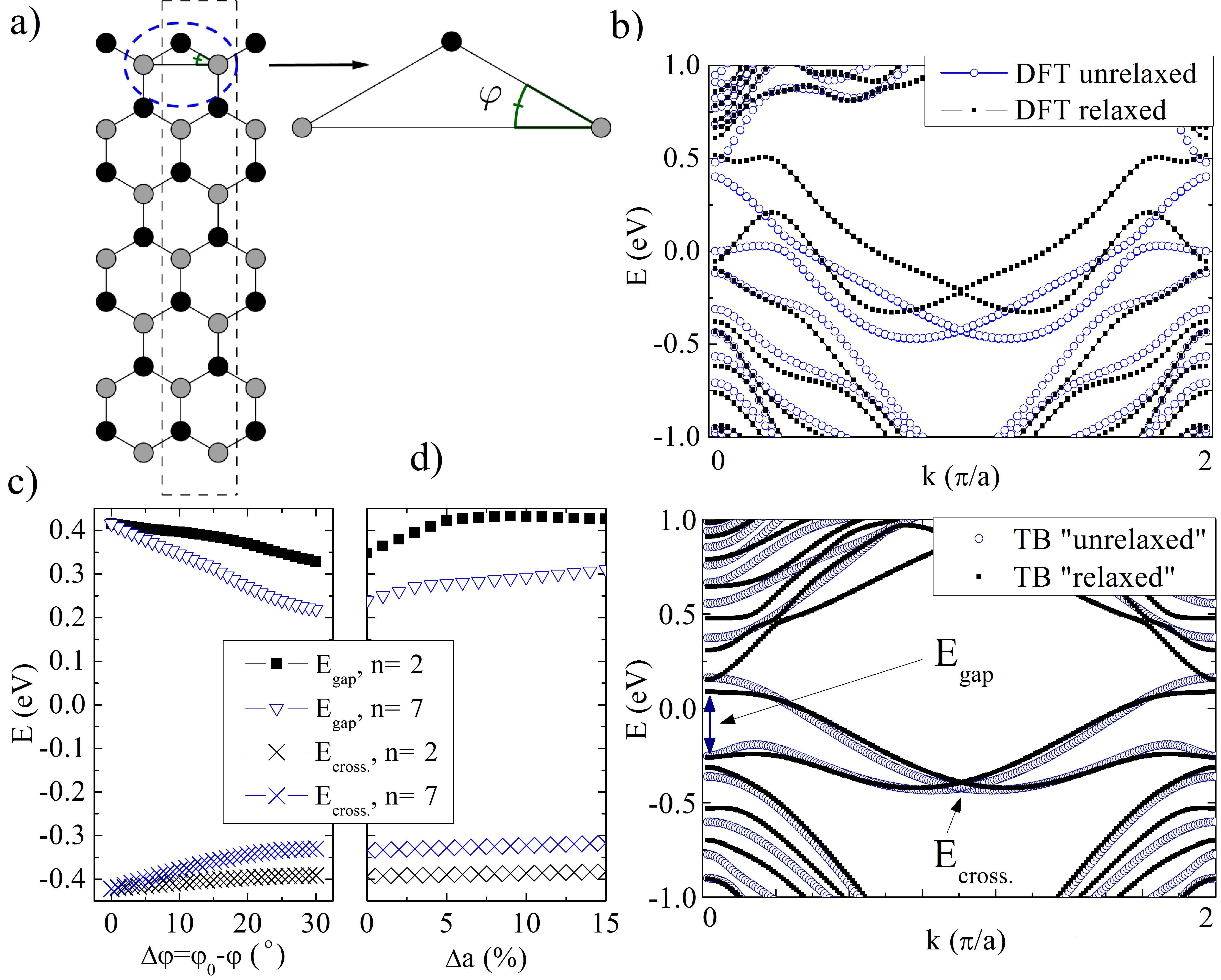}
\caption{\label{fig:F3} (a) Zigzag ribbon with $N_{at}=14$. A unit cell is marked by a dashed rectangle. An angle $\varphi$ parametrizes a position of an edge atom after relaxation. (b) The energy band structure of relaxed and unrelaxed zigzag ribbon from DFT method. (c) On the left: the energy band gap ($E_{\textrm{gap}}$) and the position of edge states crossing ($E_{\textrm{cross.}}$) as a function of angle deviation from original value, $\Delta\varphi=\varphi_{0}-\varphi$. Two values of $n=2$ and $n=7$ parametrize the variation of Slater-Koster integrals (see text) due to the bond shortening. On the right: Analogous analysis in the case of variation of a lattice constant $\Delta a$ for $\Delta\varphi=25^{\circ}$. (d) The band structures from TB method of relaxed and unrelaxed zigzag ribbon. Relaxed structure was modeled for $\Delta \phi=25^{\circ}$ and $\Delta a=0$\%. }
\end{figure}

Geometry relaxation of finite fragments of ideal honeycomb lattice is a natural process related to the stabilization of structures that leads to minimization of a total energy of the system. We investigate a band structure of geometry--optimized zigzag ribbon with $N_{at}=14$, see Fig. \ref{fig:F3}(a), performed in DFT abinit software and compare it with TB model with the modified edge hopping integrals. Relaxation affects mostly the edge lattice sites, which have only two neighbors, and the ideal edge hexagons are deformed. Edge atoms move towards a center of a hexagon which decreases their relative distance to two nearest-neighbors (by $0.04$\AA) and changes an angle $\varphi$ between atomic bonds from $30.00^{\circ}$ to $28.41^{\circ}$, see Fig. \ref{fig:F3}(a). The effect of this shift has an influence on energy band structure in a vicinity of the Fermi energy shown in Fig. \ref{fig:F3}(b) for DFT results. Smooth edge states crossing the energy gap become slightly rippled ones, but their crossing remains. Thus, we do not expect a destruction of topological properties of the system due to relaxation. We note here that small anti-crossings of edge states after geometry relaxation for narrower ribbons occur but situation shown in Fig. \ref{fig:F3}(b) is more typical for wider ribbons.  

Next we model a relaxation effect within TB method by modifying angles and hopping parameters between edge atoms and their neighbors. Hopping parameters $V_{ij}$ are changed according to Harrison theory \cite{Harrison}, $V_{ij}=V^{0}_{ij} (d/d_{0})^{-n}$, where $V^{0}_{ij}$ are the original values of Slater-Koster parameter, $d_{0}$ and $d$ are original and modified bond lengths, respectively. We take $n$ as a parameter which is responsible for the strength of $V_{ij}$ modification due to bond shortening. The effect of edge modification on the band gap ($E_{\textrm{gap}}$) and an edge state crossing energy ($E_{\textrm{cross.}}$) is presented in Fig. \ref{fig:F3}(c), where the angle $\phi$ (at the edges) and the lattice constant $a$ (in the whole system) changes are studied. For all possible positive $n$ and different $\theta$ the band gap decreases but remains open. However, an increase of a lattice constant $a$ leads to the wider band gap. Both processes increase an energy of edge state crossing, but this behavior in TB model is an order of magnitude smaller than in DFT calculations. We note that within the edge modified TB model, Fig. \ref{fig:F3}(d), we were not able to change significantly a dispersion of edge states, as seen in DFT result in Fig. \ref{fig:F3}(b). This suggests that mechanism of edge relaxation needs more sophisticated TB parametrization.

We have also investigated an effect of a substrate on a ribbon band structure. We have considered a system which consists of two infinite bilayers, where periodic boundary conditions were imposed on a bottom one. In this case, two double degenerated branches of edges states split due to the inversion symmetry breaking (not shown here). Therefore, the substrate has similar effect to an external perpendicular electric field when two sublattices of honeycomb crystal become inequivalent \cite{1Chen}. Besides the degeneracy removal, no other significant effects were noticed.

\section{Transport properties}

\begin{figure}
\includegraphics[width=0.50\textwidth]{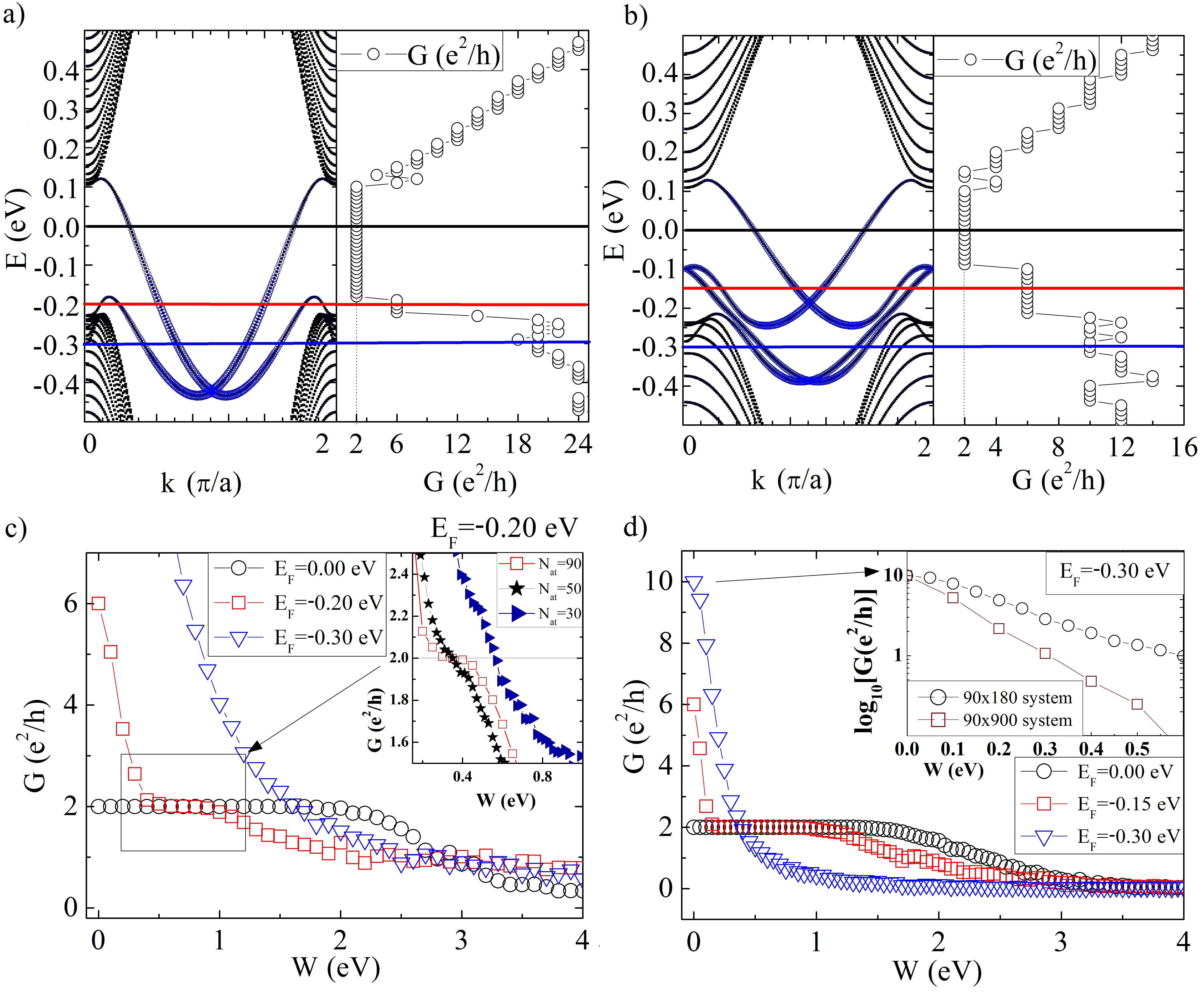}
\caption{\label{fig:F4} (a) and (b) The energy band structures (left) and transmission at a corresponding energy (right) for (a) zigzag and (b) armchair ribbons in a system without disorder. (c) and (d) The conductance of (c) zigzag and (d) armchair ribbons as a function of disorder strength $W$ for three values of Fermi energies marked on (a) and (b) by horizontal color lines. In the inset in (c) we present results for different ribbon widths, $N_{at}=30$, $50$ and $90$ atoms, keeping aspect ratio of the system constant (N$_{\textrm{width}}$=2N$_{\textrm{at.}}$). The inset in (d) compares a decrease of the conductance as a result of localization of states for energy below the top of the valence band for two different system lengths, with $180$ and $900$ atoms.} 
\end{figure}

The topological nature of the edge states ensures a protection against backscattering of transported electrons. This is related to the helical edge channels which results in a spin-momentum locking. Electrons with a given spin traveling in one direction cannot be backscattered unless the spin-flip process occurs. As it was shown in a previous Section, edge states dispersion in bismuth ribbons is not linear as in a simple TI model \cite{KanePRL}, and an electrons velocity does not have a uniquely defined direction within the entire Brillouin zone. As the edge states dispersion allows to change a movement direction, it is tempting to verify whether the backscattering in this case is still forbidden. 

Figures \ref{fig:F4}(a) and (b) present the band structures and the corresponding conductance $G$ at the given Fermi energy for zigzag and armchair ribbons with $N_{at}=90$ atoms and in the absence of disorder. We consider three characteristic regimes of the Fermi energy position: within the band gap region with the edge states crossing the Fermi energy twice and six times, and below the energy gap, where a contribution to transport from bulk states occurs, see solid lines in Fig. \ref{fig:F4}(a) and (b). All states in the band structure are Kramers degenerate. When the sample is biased, each edge state results in a single contribution to the transport, increasing the conductance by $e^{2}/h$. The conductance within the first regime in the band gap region at the Fermi energy represented by black lines in Fig. \ref{fig:F4}(a -- b) is equal to 2$e^{2}/h$. Then it increases to 6$e^{2}/h$ when the Fermi level is lowered to the point where a double crossing with the other branch of edge states occurs (red line). When the Fermi energy crosses more bulk states, they start to contribute to transport and one can observe almost monotonic increase of the conductance. Deviations from this trend are seen for some energies and are attributed to unimportant size effects in the studied system. 

For three characteristic regimes we check how disorder influences the conductance of the system. We study samples with scattering region composed of $90\times 180$ atoms with semi-infinite pure leads attached at both ends. Anderson type of disorder is introduced by adding a random on-site potential on lattice sites. When the Fermi level, represented by black lines in Fig. \ref{fig:F4}(a -- b) crosses one pair of edge states, transport is topologically protected and only very strong disorder (above $W\approx2$ eV) starts to localize the states. This results in the decreasing conductance, see Fig. \ref{fig:F4}(c) for zigzag and (d) for armchair ribbons. In zigzag ribbon, when the Fermi energy starts to cross the lower branch of edge states (a red line in Fig. \ref{fig:F4}(b)), we observe their fast localization and in a consequence a decrease of the conductance from 6 to 2$e^{2}/h$, marked with red squares in Fig. \ref{fig:F4}(c). We explain this behavior as a scattering between two states from a lower branch of edge states near boundaries of the Brillouin zone. Then, for disorder values between $0.4$ and $0.9$ eV, the conductance plateau is observed with 2$e^{2}/h$, which is attributed to topologically protected transport through states from an upper branch of edge states. The inset shows that this occurs only for sufficiently wide samples. Further lowering of the Fermi energy to the point where an overlap with bulk states is observed causes an edge states mixing and results in the diminished conductance with respect to disorder strength.  Similar effects are observed for armchair ribbon. When the upper branch of edge states starts to overlap with the lower one (the second regime), the conductance drops from 6 to 2$e^{2}/h$, even more rapidly than in zigzag ribbon. Interestingly, transport here is more stable comparing to the zigzag case. The plateau is observed for $W \approx 0.2$ eV  to $W \approx 1$ eV, however this depends on sample length (see the inset in \ref{fig:F4}(d)). When the edge states start to overlap with the bulk states, the quantized conductance is destroyed just as in zigzag case. We also note that when the ribbon is within a trivial insulator regime ($\lambda=0.8$ eV), we do not observe any signs of disorder-induced topological Anderson insulator phase\cite{11Li, 13Groth}. We studied also the effect of the substrate within TI regime. We have found that the splitting of states has no effect on the transport properties.

\begin{figure}
\includegraphics[width=0.50\textwidth]{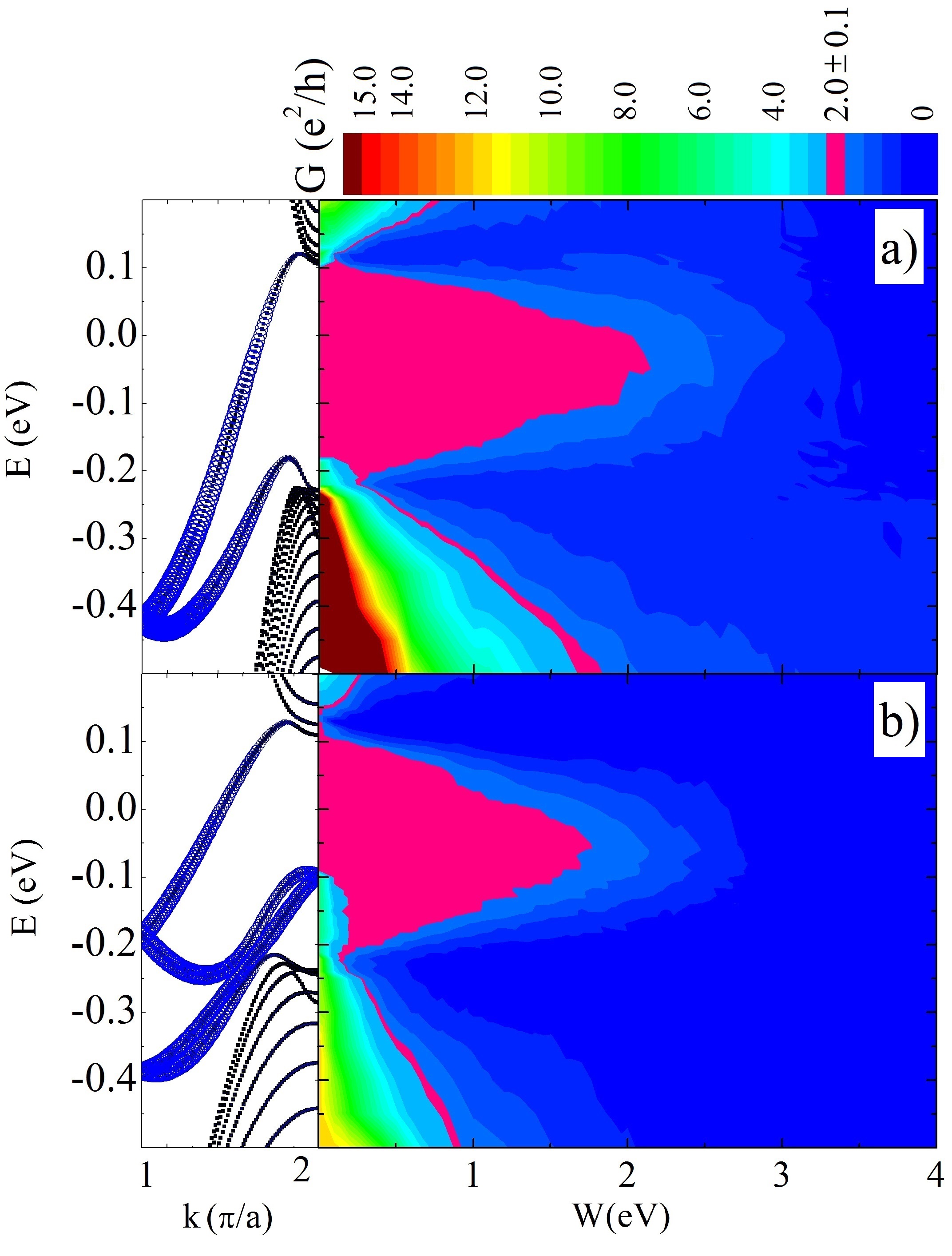}
\caption{\label{fig:F5} A map of the conductance $G$ as a function of disorder strength $W$ and  the Fermi energy $E$ for (a) zigzag and (b) armchair-type ribbons with $90\times 180$ atoms. On the left we show corresponding energy band structures for ribbons with $N_{\textrm{at}}=90$. A region with topologically protected transport with $G=2e^{2}/h$ is clearly visible.}
\end{figure}
The results of disorder effects on the conductance are summarized on maps shown in Fig. \ref{fig:F5}, (a) and (b) for zigzag and armchair ribbons, respectively. In both cases, a large region within the energy gap with the quantized conductance $G=2e^{2}/h$ is seen. In armchair case within an energy region $E=(-0.1,-0.2)$, one can observe transport with the quantized conductance $G=2e^{2}/h$ even when there are lower and upper edge state branches. Due to the short distance in reciprocal space between states from the lower branch, the scattering within them is possible, which is not true for states from the upper branch. Thus, we conclude that a value of disorder strength necessary to break the ideally quantized conductance $G=2e^{2}/h$ is affected by a distance between states in $k$--space, and also an energetic distance between the bulk and the edge states. Thus, this value does not depend on the level of localization of wavefunction at the edge, see the localization of edge states in Fig. \ref{fig:F2} (c) and (f). The conductance through the bulk states in zigzag case is higher (dark red color in Fig. \ref{fig:F5} (a)), which is related to larger density of bulk states for this kind of edge termination. 

\section{Conclusions}
In summary, we have studied electronic and transport properties of bismuth (111) bilayer and ribbons in a context of stability of their topological properties against different perturbations. We investigated the changes in energy band structures due to variations of SOC, a geometry relaxation effect and an interaction with a substrate. We studied also the Anderson-type disorder effects on transport properties of this system. We varied SOC parameter from  initial value $\lambda > 1.5$ eV and we have shown that the system transforms into a trivial insulator for $\lambda < 0.982$ eV. In bismuth (111) bilayer this is associated with an inversion of bands in the energy band structure. In ribbons a dispersion of edge states flattens, and as a consequence  a connection between valence and conduction bands by these states is destroyed. A change of dispersion of edge states associated with geometry relaxation has no effect on their topological nature. The effect of interaction with a substrate is similar to that of external perpendicular electric field and we did not study it here in detail. 

In order to verify the topological protection of edge states against backscattering, we examined the transport properties in a presence of the Anderson--type disorder. We have shown a regime with the quantized conductance unaffected by a weak disorder, where the edge states have quasi-linear dispersion within the energy band gap. We also note that within this regime no effect of an interaction with a substrate can be noticed. For energies, where an edge state  crosses the given Fermi energy twice (zigzag ribbon case), or there are two branches of edge states (armchair ribbon case), a scattering is possible between channels lying close in $k$--space. In this case, the transport through the edge channels localized in a different part of the Brillouin zone was still unaffected by a weak disorder. When the Fermi energy overlaps with the bulk states energies, no topological protection has been noticed. We have also verified that the TAI phase do not appear in bismuth when SOC was decreased changing the system into trivial insulator. 

{\it Acknowledgments}. The authors acknowledge a partial financial support from National Science Center (NCN), Poland, grant Sonata No. 2013/11/D/ST3/02703. Our calculations were performed in Wroc\l'aw Center for Networking and Supercomputing. We also acknowledge the assistance M. Brzezi\'nska and P. Scharoch in editing the manuscript.

\bibliography{apssamp}

\begin{thebibliography}{53}
\expandafter\ifx\csname natexlab\endcsname\relax\def\natexlab#1{#1}\fi
\expandafter\ifx\csname bibnamefont\endcsname\relax
  \def\bibnamefont#1{#1}\fi
\expandafter\ifx\csname bibfnamefont\endcsname\relax
  \def\bibfnamefont#1{#1}\fi
\expandafter\ifx\csname citenamefont\endcsname\relax
  \def\citenamefont#1{#1}\fi
\expandafter\ifx\csname url\endcsname\relax
  \def\url#1{\texttt{#1}}\fi
\expandafter\ifx\csname urlprefix\endcsname\relax\def\urlprefix{URL }\fi
\providecommand{\bibinfo}[2]{#2}
\providecommand{\eprint}[2][]{\url{#2}}

\bibitem[{\citenamefont{Moore}(2010)}]{4Moore}
\bibinfo{author}{\bibfnamefont{J.}~\bibnamefont{Moore}},
  \bibinfo{journal}{Nature} \textbf{\bibinfo{volume}{464}},
  \bibinfo{pages}{194} (\bibinfo{year}{2010}),
  \urlprefix\url{http://dx.doi.org/10.1038/nature08916}.

\bibitem[{\citenamefont{Hasan and Kane}(2010)}]{5Hasan}
\bibinfo{author}{\bibfnamefont{M.~Z.} \bibnamefont{Hasan}} \bibnamefont{and}
  \bibinfo{author}{\bibfnamefont{C.~L.} \bibnamefont{Kane}},
  \bibinfo{journal}{Rev. Mod. Phys.} \textbf{\bibinfo{volume}{82}},
  \bibinfo{pages}{3045} (\bibinfo{year}{2010}),
  \urlprefix\url{http://link.aps.org/doi/10.1103/RevModPhys.82.3045}.

\bibitem[{\citenamefont{Qi and Zhang}(2011)}]{6Qi}
\bibinfo{author}{\bibfnamefont{X.-L.} \bibnamefont{Qi}} \bibnamefont{and}
  \bibinfo{author}{\bibfnamefont{S.-C.} \bibnamefont{Zhang}},
  \bibinfo{journal}{Rev. Mod. Phys.} \textbf{\bibinfo{volume}{83}},
  \bibinfo{pages}{1057} (\bibinfo{year}{2011}),
  \urlprefix\url{http://link.aps.org/doi/10.1103/RevModPhys.83.1057}.

\bibitem[{\citenamefont{Kane and Mele}(2005)}]{KanePRL}
\bibinfo{author}{\bibfnamefont{C.~L.} \bibnamefont{Kane}} \bibnamefont{and}
  \bibinfo{author}{\bibfnamefont{E.~J.} \bibnamefont{Mele}},
  \bibinfo{journal}{Phys. Rev. Lett.} \textbf{\bibinfo{volume}{95}},
  \bibinfo{pages}{226801} (\bibinfo{year}{2005}),
  \urlprefix\url{http://link.aps.org/doi/10.1103/PhysRevLett.95.226801}.

\bibitem[{\citenamefont{K{\"o}nig et~al.}(2007)\citenamefont{K{\"o}nig,
  Wiedmann, Br{\"u}ne, Roth, Buhmann, Molenkamp, Qi, and Zhang}}]{3Konig}
\bibinfo{author}{\bibfnamefont{M.}~\bibnamefont{K{\"o}nig}},
  \bibinfo{author}{\bibfnamefont{S.}~\bibnamefont{Wiedmann}},
  \bibinfo{author}{\bibfnamefont{C.}~\bibnamefont{Br{\"u}ne}},
  \bibinfo{author}{\bibfnamefont{A.}~\bibnamefont{Roth}},
  \bibinfo{author}{\bibfnamefont{H.}~\bibnamefont{Buhmann}},
  \bibinfo{author}{\bibfnamefont{L.~W.} \bibnamefont{Molenkamp}},
  \bibinfo{author}{\bibfnamefont{X.-L.} \bibnamefont{Qi}}, \bibnamefont{and}
  \bibinfo{author}{\bibfnamefont{S.-C.} \bibnamefont{Zhang}},
  \bibinfo{journal}{Science} \textbf{\bibinfo{volume}{318}},
  \bibinfo{pages}{766} (\bibinfo{year}{2007}),
  \urlprefix\url{http://science.sciencemag.org/content/318/5851/766}.

\bibitem[{\citenamefont{Knez et~al.}(2011)\citenamefont{Knez, Du, and
  Sullivan}}]{3Knez}
\bibinfo{author}{\bibfnamefont{I.}~\bibnamefont{Knez}},
  \bibinfo{author}{\bibfnamefont{R.-R.} \bibnamefont{Du}}, \bibnamefont{and}
  \bibinfo{author}{\bibfnamefont{G.}~\bibnamefont{Sullivan}},
  \bibinfo{journal}{Phys. Rev. Lett.} \textbf{\bibinfo{volume}{107}},
  \bibinfo{pages}{136603} (\bibinfo{year}{2011}),
  \urlprefix\url{http://link.aps.org/doi/10.1103/PhysRevLett.107.136603}.

\bibitem[{\citenamefont{Li et~al.}(2016)\citenamefont{Li, Huang, Lv, Zhang,
  Yang, Zhang, Chen, Yao, Zhou, Lu et~al.}}]{3Li}
\bibinfo{author}{\bibfnamefont{X.-B.} \bibnamefont{Li}},
  \bibinfo{author}{\bibfnamefont{W.-K.} \bibnamefont{Huang}},
  \bibinfo{author}{\bibfnamefont{Y.-Y.} \bibnamefont{Lv}},
  \bibinfo{author}{\bibfnamefont{K.-W.} \bibnamefont{Zhang}},
  \bibinfo{author}{\bibfnamefont{C.-L.} \bibnamefont{Yang}},
  \bibinfo{author}{\bibfnamefont{B.-B.} \bibnamefont{Zhang}},
  \bibinfo{author}{\bibfnamefont{Y.~B.} \bibnamefont{Chen}},
  \bibinfo{author}{\bibfnamefont{S.-H.} \bibnamefont{Yao}},
  \bibinfo{author}{\bibfnamefont{J.}~\bibnamefont{Zhou}},
  \bibinfo{author}{\bibfnamefont{M.-H.} \bibnamefont{Lu}},
  \bibnamefont{et~al.}, \bibinfo{journal}{Phys. Rev. Lett.}
  \textbf{\bibinfo{volume}{116}}, \bibinfo{pages}{176803}
  (\bibinfo{year}{2016}),
  \urlprefix\url{http://link.aps.org/doi/10.1103/PhysRevLett.116.176803}.

\bibitem[{\citenamefont{Wu et~al.}(2016)\citenamefont{Wu, Ma, Nie, Zhao, Huang,
  Yin, Fu, Richard, Chen, Fang et~al.}}]{3Wu}
\bibinfo{author}{\bibfnamefont{R.}~\bibnamefont{Wu}},
  \bibinfo{author}{\bibfnamefont{J.-Z.} \bibnamefont{Ma}},
  \bibinfo{author}{\bibfnamefont{S.-M.} \bibnamefont{Nie}},
  \bibinfo{author}{\bibfnamefont{L.-X.} \bibnamefont{Zhao}},
  \bibinfo{author}{\bibfnamefont{X.}~\bibnamefont{Huang}},
  \bibinfo{author}{\bibfnamefont{J.-X.} \bibnamefont{Yin}},
  \bibinfo{author}{\bibfnamefont{B.-B.} \bibnamefont{Fu}},
  \bibinfo{author}{\bibfnamefont{P.}~\bibnamefont{Richard}},
  \bibinfo{author}{\bibfnamefont{G.-F.} \bibnamefont{Chen}},
  \bibinfo{author}{\bibfnamefont{Z.}~\bibnamefont{Fang}}, \bibnamefont{et~al.},
  \bibinfo{journal}{Phys. Rev. X} \textbf{\bibinfo{volume}{6}},
  \bibinfo{pages}{021017} (\bibinfo{year}{2016}),
  \urlprefix\url{http://link.aps.org/doi/10.1103/PhysRevX.6.021017}.

\bibitem[{\citenamefont{Lu et~al.}(2015)\citenamefont{Lu, Xu, Zeng, Yao, Shen,
  Yang, Luo, Pan, Wu, Das et~al.}}]{3Lu}
\bibinfo{author}{\bibfnamefont{Y.}~\bibnamefont{Lu}},
  \bibinfo{author}{\bibfnamefont{W.}~\bibnamefont{Xu}},
  \bibinfo{author}{\bibfnamefont{M.}~\bibnamefont{Zeng}},
  \bibinfo{author}{\bibfnamefont{G.}~\bibnamefont{Yao}},
  \bibinfo{author}{\bibfnamefont{L.}~\bibnamefont{Shen}},
  \bibinfo{author}{\bibfnamefont{M.}~\bibnamefont{Yang}},
  \bibinfo{author}{\bibfnamefont{Z.}~\bibnamefont{Luo}},
  \bibinfo{author}{\bibfnamefont{F.}~\bibnamefont{Pan}},
  \bibinfo{author}{\bibfnamefont{K.}~\bibnamefont{Wu}},
  \bibinfo{author}{\bibfnamefont{T.}~\bibnamefont{Das}}, \bibnamefont{et~al.},
  \bibinfo{journal}{Nano Letters} \textbf{\bibinfo{volume}{15}},
  \bibinfo{pages}{80} (\bibinfo{year}{2015}),
  \urlprefix\url{http://dx.doi.org/10.1021/nl502997v}.

\bibitem[{\citenamefont{Yang et~al.}(2012)\citenamefont{Yang, Miao, Wang, Yao,
  Zhu, Song, Wang, Xu, Fedorov, Sun et~al.}}]{1Yang}
\bibinfo{author}{\bibfnamefont{F.}~\bibnamefont{Yang}},
  \bibinfo{author}{\bibfnamefont{L.}~\bibnamefont{Miao}},
  \bibinfo{author}{\bibfnamefont{Z.~F.} \bibnamefont{Wang}},
  \bibinfo{author}{\bibfnamefont{M.-Y.} \bibnamefont{Yao}},
  \bibinfo{author}{\bibfnamefont{F.}~\bibnamefont{Zhu}},
  \bibinfo{author}{\bibfnamefont{Y.~R.} \bibnamefont{Song}},
  \bibinfo{author}{\bibfnamefont{M.-X.} \bibnamefont{Wang}},
  \bibinfo{author}{\bibfnamefont{J.-P.} \bibnamefont{Xu}},
  \bibinfo{author}{\bibfnamefont{A.~V.} \bibnamefont{Fedorov}},
  \bibinfo{author}{\bibfnamefont{Z.}~\bibnamefont{Sun}}, \bibnamefont{et~al.},
  \bibinfo{journal}{Phys. Rev. Lett.} \textbf{\bibinfo{volume}{109}},
  \bibinfo{pages}{016801} (\bibinfo{year}{2012}),
  \urlprefix\url{http://link.aps.org/doi/10.1103/PhysRevLett.109.016801}.

\bibitem[{\citenamefont{Sabater et~al.}(2013)\citenamefont{Sabater,
  Gos\'albez-Mart\'{\i}nez, Fern\'andez-Rossier, Rodrigo, Untiedt, and
  Palacios}}]{1Sabater}
\bibinfo{author}{\bibfnamefont{C.}~\bibnamefont{Sabater}},
  \bibinfo{author}{\bibfnamefont{D.}~\bibnamefont{Gos\'albez-Mart\'{\i}nez}},
  \bibinfo{author}{\bibfnamefont{J.}~\bibnamefont{Fern\'andez-Rossier}},
  \bibinfo{author}{\bibfnamefont{J.~G.} \bibnamefont{Rodrigo}},
  \bibinfo{author}{\bibfnamefont{C.}~\bibnamefont{Untiedt}}, \bibnamefont{and}
  \bibinfo{author}{\bibfnamefont{J.~J.} \bibnamefont{Palacios}},
  \bibinfo{journal}{Phys. Rev. Lett.} \textbf{\bibinfo{volume}{110}},
  \bibinfo{pages}{176802} (\bibinfo{year}{2013}),
  \urlprefix\url{http://link.aps.org/doi/10.1103/PhysRevLett.110.176802}.

\bibitem[{\citenamefont{Drozdov et~al.}(2014)\citenamefont{Drozdov,
  Alexandradinata, Jeon, Nadj-Perge, Ji, Cava, Bernevig, and
  Yazdani}}]{1Drozdov}
\bibinfo{author}{\bibfnamefont{I.~K.} \bibnamefont{Drozdov}},
  \bibinfo{author}{\bibfnamefont{A.}~\bibnamefont{Alexandradinata}},
  \bibinfo{author}{\bibfnamefont{S.}~\bibnamefont{Jeon}},
  \bibinfo{author}{\bibfnamefont{S.}~\bibnamefont{Nadj-Perge}},
  \bibinfo{author}{\bibfnamefont{H.}~\bibnamefont{Ji}},
  \bibinfo{author}{\bibfnamefont{R.~J.} \bibnamefont{Cava}},
  \bibinfo{author}{\bibfnamefont{B.~A.} \bibnamefont{Bernevig}},
  \bibnamefont{and} \bibinfo{author}{\bibfnamefont{A.}~\bibnamefont{Yazdani}},
  \bibinfo{journal}{Nature Physics} \textbf{\bibinfo{volume}{10}},
  \bibinfo{pages}{664} (\bibinfo{year}{2014}),
  \urlprefix\url{http://dx.doi.org/10.1038/nphys3048}.

\bibitem[{\citenamefont{Ren et~al.}(2016)\citenamefont{Ren, Qiao, and
  Niu}}]{1Ren}
\bibinfo{author}{\bibfnamefont{Y.}~\bibnamefont{Ren}},
  \bibinfo{author}{\bibfnamefont{Z.}~\bibnamefont{Qiao}}, \bibnamefont{and}
  \bibinfo{author}{\bibfnamefont{Q.}~\bibnamefont{Niu}},
  \bibinfo{journal}{Reports on Progress in Physics}
  \textbf{\bibinfo{volume}{79}}, \bibinfo{pages}{066501}
  (\bibinfo{year}{2016}),
  \urlprefix\url{http://stacks.iop.org/0034-4885/79/i=6/a=066501}.

\bibitem[{\citenamefont{Ando}(2013)}]{1Ando}
\bibinfo{author}{\bibfnamefont{Y.}~\bibnamefont{Ando}},
  \bibinfo{journal}{Journal of the Physical Society of Japan}
  \textbf{\bibinfo{volume}{82}}, \bibinfo{pages}{102001}
  (\bibinfo{year}{2013}),
  \urlprefix\url{http://dx.doi.org/10.7566/JPSJ.82.102001}.

\bibitem[{\citenamefont{Zhang et~al.}(2014)\citenamefont{Zhang, Shen, An, Sun,
  Xie, Chang, and Li}}]{ZhangPRB90}
\bibinfo{author}{\bibfnamefont{Y.-Y.} \bibnamefont{Zhang}},
  \bibinfo{author}{\bibfnamefont{M.}~\bibnamefont{Shen}},
  \bibinfo{author}{\bibfnamefont{X.-T.} \bibnamefont{An}},
  \bibinfo{author}{\bibfnamefont{Q.-F.} \bibnamefont{Sun}},
  \bibinfo{author}{\bibfnamefont{X.-C.} \bibnamefont{Xie}},
  \bibinfo{author}{\bibfnamefont{K.}~\bibnamefont{Chang}}, \bibnamefont{and}
  \bibinfo{author}{\bibfnamefont{S.-S.} \bibnamefont{Li}},
  \bibinfo{journal}{Phys. Rev. B} \textbf{\bibinfo{volume}{90}},
  \bibinfo{pages}{054205} (\bibinfo{year}{2014}),
  \urlprefix\url{http://link.aps.org/doi/10.1103/PhysRevB.90.054205}.

\bibitem[{\citenamefont{Baum et~al.}(2015)\citenamefont{Baum, Posske, Fulga,
  Trauzettel, and Stern}}]{BaumPRL114}
\bibinfo{author}{\bibfnamefont{Y.}~\bibnamefont{Baum}},
  \bibinfo{author}{\bibfnamefont{T.}~\bibnamefont{Posske}},
  \bibinfo{author}{\bibfnamefont{I.~C.} \bibnamefont{Fulga}},
  \bibinfo{author}{\bibfnamefont{B.}~\bibnamefont{Trauzettel}},
  \bibnamefont{and} \bibinfo{author}{\bibfnamefont{A.}~\bibnamefont{Stern}},
  \bibinfo{journal}{Phys. Rev. Lett.} \textbf{\bibinfo{volume}{114}},
  \bibinfo{pages}{136801} (\bibinfo{year}{2015}),
  \urlprefix\url{http://link.aps.org/doi/10.1103/PhysRevLett.114.136801}.

\bibitem[{\citenamefont{Saha and Garate}(2014)}]{SahaPRB90}
\bibinfo{author}{\bibfnamefont{K.}~\bibnamefont{Saha}} \bibnamefont{and}
  \bibinfo{author}{\bibfnamefont{I.}~\bibnamefont{Garate}},
  \bibinfo{journal}{Phys. Rev. B} \textbf{\bibinfo{volume}{90}},
  \bibinfo{pages}{245418} (\bibinfo{year}{2014}),
  \urlprefix\url{http://link.aps.org/doi/10.1103/PhysRevB.90.245418}.

\bibitem[{\citenamefont{Murakami}(2006)}]{MurakamiPRL97}
\bibinfo{author}{\bibfnamefont{S.}~\bibnamefont{Murakami}},
  \bibinfo{journal}{Phys. Rev. Lett.} \textbf{\bibinfo{volume}{97}},
  \bibinfo{pages}{236805} (\bibinfo{year}{2006}),
  \urlprefix\url{http://link.aps.org/doi/10.1103/PhysRevLett.97.236805}.

\bibitem[{\citenamefont{Zhou et~al.}(2014)\citenamefont{Zhou, Ming, Liu, Wang,
  Li, and Liu}}]{1Zhou}
\bibinfo{author}{\bibfnamefont{M.}~\bibnamefont{Zhou}},
  \bibinfo{author}{\bibfnamefont{W.}~\bibnamefont{Ming}},
  \bibinfo{author}{\bibfnamefont{Z.}~\bibnamefont{Liu}},
  \bibinfo{author}{\bibfnamefont{Z.}~\bibnamefont{Wang}},
  \bibinfo{author}{\bibfnamefont{P.}~\bibnamefont{Li}}, \bibnamefont{and}
  \bibinfo{author}{\bibfnamefont{F.}~\bibnamefont{Liu}},
  \bibinfo{journal}{Proc. Natl Acad. Sci.} \textbf{\bibinfo{volume}{111}},
  \bibinfo{pages}{14378} (\bibinfo{year}{2014}),
  \urlprefix\url{http://dx.doi.org/10.1073/pnas.1409701111}.

\bibitem[{\citenamefont{Chen et~al.}(2013)\citenamefont{Chen, Wang, and
  Liu}}]{1Chen}
\bibinfo{author}{\bibfnamefont{L.}~\bibnamefont{Chen}},
  \bibinfo{author}{\bibfnamefont{Z.~F.} \bibnamefont{Wang}}, \bibnamefont{and}
  \bibinfo{author}{\bibfnamefont{F.}~\bibnamefont{Liu}},
  \bibinfo{journal}{Phys. Rev. B} \textbf{\bibinfo{volume}{87}},
  \bibinfo{pages}{235420} (\bibinfo{year}{2013}),
  \urlprefix\url{http://dx.doi.org/10.1103/PhysRevB.87.235420}.

\bibitem[{\citenamefont{Huang et~al.}(2013)\citenamefont{Huang, Chuang, Hsu,
  Liu, Chang, Lin, and Bansil}}]{HuangPRB88}
\bibinfo{author}{\bibfnamefont{Z.-Q.} \bibnamefont{Huang}},
  \bibinfo{author}{\bibfnamefont{F.-C.} \bibnamefont{Chuang}},
  \bibinfo{author}{\bibfnamefont{C.-H.} \bibnamefont{Hsu}},
  \bibinfo{author}{\bibfnamefont{Y.-T.} \bibnamefont{Liu}},
  \bibinfo{author}{\bibfnamefont{H.-R.} \bibnamefont{Chang}},
  \bibinfo{author}{\bibfnamefont{H.}~\bibnamefont{Lin}}, \bibnamefont{and}
  \bibinfo{author}{\bibfnamefont{A.}~\bibnamefont{Bansil}},
  \bibinfo{journal}{Phys. Rev. B} \textbf{\bibinfo{volume}{88}},
  \bibinfo{pages}{165301} (\bibinfo{year}{2013}),
  \urlprefix\url{http://link.aps.org/doi/10.1103/PhysRevB.88.165301}.

\bibitem[{\citenamefont{Bernevig et~al.}(2006)\citenamefont{Bernevig, Hughes,
  and Zhang}}]{BernevigSCI}
\bibinfo{author}{\bibfnamefont{B.~A.} \bibnamefont{Bernevig}},
  \bibinfo{author}{\bibfnamefont{T.~L.} \bibnamefont{Hughes}},
  \bibnamefont{and} \bibinfo{author}{\bibfnamefont{S.-C.} \bibnamefont{Zhang}},
  \bibinfo{journal}{Science} \textbf{\bibinfo{volume}{314}},
  \bibinfo{pages}{1757} (\bibinfo{year}{2006}),
  \urlprefix\url{http://science.sciencemag.org/content/314/5806/1757}.

\bibitem[{\citenamefont{Wada et~al.}(2011)\citenamefont{Wada, Murakami,
  Freimuth, and Bihlmayer}}]{WadaPRB83}
\bibinfo{author}{\bibfnamefont{M.}~\bibnamefont{Wada}},
  \bibinfo{author}{\bibfnamefont{S.}~\bibnamefont{Murakami}},
  \bibinfo{author}{\bibfnamefont{F.}~\bibnamefont{Freimuth}}, \bibnamefont{and}
  \bibinfo{author}{\bibfnamefont{G.}~\bibnamefont{Bihlmayer}},
  \bibinfo{journal}{Phys. Rev. B} \textbf{\bibinfo{volume}{83}},
  \bibinfo{pages}{121310} (\bibinfo{year}{2011}),
  \urlprefix\url{http://link.aps.org/doi/10.1103/PhysRevB.83.121310}.

\bibitem[{\citenamefont{Chen et~al.}(2014)\citenamefont{Chen, Cui, Zhang, Wang,
  Liu, and Wang}}]{ChenRSC}
\bibinfo{author}{\bibfnamefont{L.}~\bibnamefont{Chen}},
  \bibinfo{author}{\bibfnamefont{G.}~\bibnamefont{Cui}},
  \bibinfo{author}{\bibfnamefont{P.}~\bibnamefont{Zhang}},
  \bibinfo{author}{\bibfnamefont{X.}~\bibnamefont{Wang}},
  \bibinfo{author}{\bibfnamefont{H.}~\bibnamefont{Liu}}, \bibnamefont{and}
  \bibinfo{author}{\bibfnamefont{D.}~\bibnamefont{Wang}},
  \bibinfo{journal}{Phys.Chem.Chem.Phys.} \textbf{\bibinfo{volume}{16}},
  \bibinfo{pages}{17206} (\bibinfo{year}{2014}),
  \urlprefix\url{http://dx.doi.org/10.1039/C4CP02213K}.

\bibitem[{\citenamefont{Jin and Jhi}(2014)}]{JinSR}
\bibinfo{author}{\bibfnamefont{K.-H.} \bibnamefont{Jin}} \bibnamefont{and}
  \bibinfo{author}{\bibfnamefont{S.-H.} \bibnamefont{Jhi}},
  \bibinfo{journal}{Sci. Rep.} \textbf{\bibinfo{volume}{5}},
  \bibinfo{pages}{8426} (\bibinfo{year}{2014}),
  \urlprefix\url{http://dx.doi.org/10.1038/srep08426}.

\bibitem[{\citenamefont{Jin and Jhi}(2016)}]{JinRSC}
\bibinfo{author}{\bibfnamefont{K.-H.} \bibnamefont{Jin}} \bibnamefont{and}
  \bibinfo{author}{\bibfnamefont{S.-H.} \bibnamefont{Jhi}},
  \bibinfo{journal}{Phys. Chem. Chem. Phys.} \textbf{\bibinfo{volume}{18}},
  \bibinfo{pages}{8637} (\bibinfo{year}{2016}),
  \urlprefix\url{http://dx.doi.org/10.1039/C5CP07963B}.

\bibitem[{\citenamefont{Ma et~al.}(2015{\natexlab{a}})\citenamefont{Ma, Dai,
  Kou, Frauenheim, and Heine}}]{MaNano}
\bibinfo{author}{\bibfnamefont{Y.}~\bibnamefont{Ma}},
  \bibinfo{author}{\bibfnamefont{Y.}~\bibnamefont{Dai}},
  \bibinfo{author}{\bibfnamefont{L.}~\bibnamefont{Kou}},
  \bibinfo{author}{\bibfnamefont{T.}~\bibnamefont{Frauenheim}},
  \bibnamefont{and} \bibinfo{author}{\bibfnamefont{T.}~\bibnamefont{Heine}},
  \bibinfo{journal}{Nano Lett.} \textbf{\bibinfo{volume}{15}},
  \bibinfo{pages}{1083} (\bibinfo{year}{2015}{\natexlab{a}}),
  \urlprefix\url{http://dx.doi.org/10.1021/nl504037u}.

\bibitem[{\citenamefont{Ma et~al.}(2015{\natexlab{b}})\citenamefont{Ma, Li,
  Kou, Yan, Niu, Dai, and Heine}}]{MaPRB91}
\bibinfo{author}{\bibfnamefont{Y.}~\bibnamefont{Ma}},
  \bibinfo{author}{\bibfnamefont{X.}~\bibnamefont{Li}},
  \bibinfo{author}{\bibfnamefont{L.}~\bibnamefont{Kou}},
  \bibinfo{author}{\bibfnamefont{B.}~\bibnamefont{Yan}},
  \bibinfo{author}{\bibfnamefont{C.}~\bibnamefont{Niu}},
  \bibinfo{author}{\bibfnamefont{Y.}~\bibnamefont{Dai}}, \bibnamefont{and}
  \bibinfo{author}{\bibfnamefont{T.}~\bibnamefont{Heine}},
  \bibinfo{journal}{Phys. Rev. B} \textbf{\bibinfo{volume}{91}},
  \bibinfo{pages}{235306} (\bibinfo{year}{2015}{\natexlab{b}}),
  \urlprefix\url{http://link.aps.org/doi/10.1103/PhysRevB.91.235306}.

\bibitem[{\citenamefont{Niu et~al.}(2015)\citenamefont{Niu, Bihlmayer, Zhang,
  Wortmann, Bl\"ugel, and Mokrousov}}]{NiuPRB91}
\bibinfo{author}{\bibfnamefont{C.}~\bibnamefont{Niu}},
  \bibinfo{author}{\bibfnamefont{G.}~\bibnamefont{Bihlmayer}},
  \bibinfo{author}{\bibfnamefont{H.}~\bibnamefont{Zhang}},
  \bibinfo{author}{\bibfnamefont{D.}~\bibnamefont{Wortmann}},
  \bibinfo{author}{\bibfnamefont{S.}~\bibnamefont{Bl\"ugel}}, \bibnamefont{and}
  \bibinfo{author}{\bibfnamefont{Y.}~\bibnamefont{Mokrousov}},
  \bibinfo{journal}{Phys. Rev. B} \textbf{\bibinfo{volume}{91}},
  \bibinfo{pages}{041303} (\bibinfo{year}{2015}),
  \urlprefix\url{http://link.aps.org/doi/10.1103/PhysRevB.91.041303}.

\bibitem[{\citenamefont{Wang et~al.}(2014)\citenamefont{Wang, Chen, and
  Liu}}]{WangNano}
\bibinfo{author}{\bibfnamefont{Z.~F.} \bibnamefont{Wang}},
  \bibinfo{author}{\bibfnamefont{L.}~\bibnamefont{Chen}}, \bibnamefont{and}
  \bibinfo{author}{\bibfnamefont{F.}~\bibnamefont{Liu}}, \bibinfo{journal}{Nano
  Lett.} \textbf{\bibinfo{volume}{14}}, \bibinfo{pages}{2879}
  (\bibinfo{year}{2014}), \urlprefix\url{http://dx.doi.org/10.1021/nl5009212}.

\bibitem[{\citenamefont{Wang et~al.}(2015)\citenamefont{Wang, Chen, Liu, Wang,
  Cui, Zhang, Zhao, and Ji}}]{WangRSC}
\bibinfo{author}{\bibfnamefont{D.}~\bibnamefont{Wang}},
  \bibinfo{author}{\bibfnamefont{L.}~\bibnamefont{Chen}},
  \bibinfo{author}{\bibfnamefont{H.}~\bibnamefont{Liu}},
  \bibinfo{author}{\bibfnamefont{X.}~\bibnamefont{Wang}},
  \bibinfo{author}{\bibfnamefont{G.}~\bibnamefont{Cui}},
  \bibinfo{author}{\bibfnamefont{P.}~\bibnamefont{Zhang}},
  \bibinfo{author}{\bibfnamefont{D.}~\bibnamefont{Zhao}}, \bibnamefont{and}
  \bibinfo{author}{\bibfnamefont{S.}~\bibnamefont{Ji}},
  \bibinfo{journal}{Phys.Chem.Chem.Phys.} \textbf{\bibinfo{volume}{17}},
  \bibinfo{pages}{3577} (\bibinfo{year}{2015}),
  \urlprefix\url{http://dx.doi.org/10.1039/C4CP04502E}.

\bibitem[{\citenamefont{Li et~al.}(2014)\citenamefont{Li, Liu, Jiang, Wang, and
  Feng}}]{LiPRB90}
\bibinfo{author}{\bibfnamefont{X.}~\bibnamefont{Li}},
  \bibinfo{author}{\bibfnamefont{H.}~\bibnamefont{Liu}},
  \bibinfo{author}{\bibfnamefont{H.}~\bibnamefont{Jiang}},
  \bibinfo{author}{\bibfnamefont{F.}~\bibnamefont{Wang}}, \bibnamefont{and}
  \bibinfo{author}{\bibfnamefont{J.}~\bibnamefont{Feng}},
  \bibinfo{journal}{Phys. Rev. B} \textbf{\bibinfo{volume}{90}},
  \bibinfo{pages}{165412} (\bibinfo{year}{2014}),
  \urlprefix\url{http://link.aps.org/doi/10.1103/PhysRevB.90.165412}.

\bibitem[{\citenamefont{Potasz and Fern\'andez-Rossier}(2015)}]{PPNanoLett}
\bibinfo{author}{\bibfnamefont{P.}~\bibnamefont{Potasz}} \bibnamefont{and}
  \bibinfo{author}{\bibfnamefont{J.}~\bibnamefont{Fern\'andez-Rossier}},
  \bibinfo{journal}{Nano Letters} \textbf{\bibinfo{volume}{15}},
  \bibinfo{pages}{5799} (\bibinfo{year}{2015}),
  \urlprefix\url{http://dx.doi.org/10.1021/acs.nanolett.5b01805}.

\bibitem[{\citenamefont{Hofmann}(2006)}]{Hofmann}
\bibinfo{author}{\bibfnamefont{P.}~\bibnamefont{Hofmann}},
  \bibinfo{journal}{Prog. Surf. Sci.} \textbf{\bibinfo{volume}{81}},
  \bibinfo{pages}{191} (\bibinfo{year}{2006}),
  \urlprefix\url{http://dx.doi.org/10.1016/j.progsurf.2006.03.001}.

\bibitem[{\citenamefont{Ohtsubo et~al.}(2013)\citenamefont{Ohtsubo, Perfetti,
  Goerbig, Fèvre, Bertran, and Taleb-Ibrahimi}}]{OhtsuboNJP}
\bibinfo{author}{\bibfnamefont{Y.}~\bibnamefont{Ohtsubo}},
  \bibinfo{author}{\bibfnamefont{L.}~\bibnamefont{Perfetti}},
  \bibinfo{author}{\bibfnamefont{M.~O.} \bibnamefont{Goerbig}},
  \bibinfo{author}{\bibfnamefont{P.~L.} \bibnamefont{Fèvre}},
  \bibinfo{author}{\bibfnamefont{F.}~\bibnamefont{Bertran}}, \bibnamefont{and}
  \bibinfo{author}{\bibfnamefont{A.}~\bibnamefont{Taleb-Ibrahimi}},
  \bibinfo{journal}{New Journal of Physics} \textbf{\bibinfo{volume}{15}},
  \bibinfo{pages}{033041} (\bibinfo{year}{2013}),
  \urlprefix\url{http://stacks.iop.org/1367-2630/15/i=3/a=033041}.

\bibitem[{\citenamefont{Liu et~al.}(2011)\citenamefont{Liu, Liu, Wu, Duan, Liu,
  and Wu}}]{LiuPRL107}
\bibinfo{author}{\bibfnamefont{Z.}~\bibnamefont{Liu}},
  \bibinfo{author}{\bibfnamefont{C.-X.} \bibnamefont{Liu}},
  \bibinfo{author}{\bibfnamefont{Y.-S.} \bibnamefont{Wu}},
  \bibinfo{author}{\bibfnamefont{W.-H.} \bibnamefont{Duan}},
  \bibinfo{author}{\bibfnamefont{F.}~\bibnamefont{Liu}}, \bibnamefont{and}
  \bibinfo{author}{\bibfnamefont{J.}~\bibnamefont{Wu}}, \bibinfo{journal}{Phys.
  Rev. Lett.} \textbf{\bibinfo{volume}{107}}, \bibinfo{pages}{136805}
  (\bibinfo{year}{2011}),
  \urlprefix\url{http://link.aps.org/doi/10.1103/PhysRevLett.107.136805}.

\bibitem[{\citenamefont{Yeom et~al.}(2016)\citenamefont{Yeom, Jin, and
  Jhi}}]{YeomPRB93}
\bibinfo{author}{\bibfnamefont{H.~W.} \bibnamefont{Yeom}},
  \bibinfo{author}{\bibfnamefont{K.-H.} \bibnamefont{Jin}}, \bibnamefont{and}
  \bibinfo{author}{\bibfnamefont{S.-H.} \bibnamefont{Jhi}},
  \bibinfo{journal}{Phys. Rev. B} \textbf{\bibinfo{volume}{93}},
  \bibinfo{pages}{075435} (\bibinfo{year}{2016}),
  \urlprefix\url{http://link.aps.org/doi/10.1103/PhysRevB.93.075435}.

\bibitem[{\citenamefont{Li et~al.}(2009)\citenamefont{Li, Chu, Jain, and
  Shen}}]{11Li}
\bibinfo{author}{\bibfnamefont{J.}~\bibnamefont{Li}},
  \bibinfo{author}{\bibfnamefont{R.-L.} \bibnamefont{Chu}},
  \bibinfo{author}{\bibfnamefont{J.~K.} \bibnamefont{Jain}}, \bibnamefont{and}
  \bibinfo{author}{\bibfnamefont{S.-Q.} \bibnamefont{Shen}},
  \bibinfo{journal}{Phys. Rev. Lett.} \textbf{\bibinfo{volume}{102}},
  \bibinfo{pages}{136806} (\bibinfo{year}{2009}),
  \urlprefix\url{http://link.aps.org/doi/10.1103/PhysRevLett.102.136806}.

\bibitem[{\citenamefont{Groth et~al.}(2009)\citenamefont{Groth, Wimmer,
  Akhmerov, Tworzyd\l{}o, and Beenakker}}]{13Groth}
\bibinfo{author}{\bibfnamefont{C.~W.} \bibnamefont{Groth}},
  \bibinfo{author}{\bibfnamefont{M.}~\bibnamefont{Wimmer}},
  \bibinfo{author}{\bibfnamefont{A.~R.} \bibnamefont{Akhmerov}},
  \bibinfo{author}{\bibfnamefont{J.}~\bibnamefont{Tworzyd\l{}o}},
  \bibnamefont{and} \bibinfo{author}{\bibfnamefont{C.~W.~J.}
  \bibnamefont{Beenakker}}, \bibinfo{journal}{Phys. Rev. Lett.}
  \textbf{\bibinfo{volume}{103}}, \bibinfo{pages}{196805}
  (\bibinfo{year}{2009}),
  \urlprefix\url{http://link.aps.org/doi/10.1103/PhysRevLett.103.196805}.

\bibitem[{\citenamefont{Gonze et~al.}(2009)\citenamefont{Gonze, Amadon,
  Anglade, Beuken, Bottin, Boulanger, Bruneval, Caliste, Caracas, Cote
  et~al.}}]{Abinit}
\bibinfo{author}{\bibfnamefont{X.}~\bibnamefont{Gonze}},
  \bibinfo{author}{\bibfnamefont{B.}~\bibnamefont{Amadon}},
  \bibinfo{author}{\bibfnamefont{P.}~\bibnamefont{Anglade}},
  \bibinfo{author}{\bibfnamefont{J.}~\bibnamefont{Beuken}},
  \bibinfo{author}{\bibfnamefont{F.}~\bibnamefont{Bottin}},
  \bibinfo{author}{\bibfnamefont{P.}~\bibnamefont{Boulanger}},
  \bibinfo{author}{\bibfnamefont{F.}~\bibnamefont{Bruneval}},
  \bibinfo{author}{\bibfnamefont{D.}~\bibnamefont{Caliste}},
  \bibinfo{author}{\bibfnamefont{R.}~\bibnamefont{Caracas}},
  \bibinfo{author}{\bibfnamefont{M.}~\bibnamefont{Cote}}, \bibnamefont{et~al.},
  \bibinfo{journal}{Comput. Phys. Commun.} \textbf{\bibinfo{volume}{180}},
  \bibinfo{pages}{2582} (\bibinfo{year}{2009}),
  \urlprefix\url{http://dx.doi.org/10.1016/j.cpc.2009.07.007}.

\bibitem[{\citenamefont{Holzwarth et~al.}(2001)\citenamefont{Holzwarth,
  Tackett, and Matthews}}]{PAW}
\bibinfo{author}{\bibfnamefont{N.}~\bibnamefont{Holzwarth}},
  \bibinfo{author}{\bibfnamefont{A.}~\bibnamefont{Tackett}}, \bibnamefont{and}
  \bibinfo{author}{\bibfnamefont{G.}~\bibnamefont{Matthews}},
  \bibinfo{journal}{Comput. Phys. Commun.} \textbf{\bibinfo{volume}{135}},
  \bibinfo{pages}{329} (\bibinfo{year}{2001}),
  \urlprefix\url{http://dx.doi.org/10.1016/S0010-4655(00)00244-7}.

\bibitem[{\citenamefont{Perdew et~al.}(1996)\citenamefont{Perdew, Burke, and
  Ernzerhof}}]{GGA}
\bibinfo{author}{\bibfnamefont{J.~P.} \bibnamefont{Perdew}},
  \bibinfo{author}{\bibfnamefont{K.}~\bibnamefont{Burke}}, \bibnamefont{and}
  \bibinfo{author}{\bibfnamefont{M.}~\bibnamefont{Ernzerhof}},
  \bibinfo{journal}{Phys. Rev. Lett.} \textbf{\bibinfo{volume}{77}},
  \bibinfo{pages}{3865} (\bibinfo{year}{1996}),
  \urlprefix\url{http://link.aps.org/doi/10.1103/PhysRevLett.77.3865}.

\bibitem[{\citenamefont{Hirahara et~al.}(2012)\citenamefont{Hirahara, Fukui,
  Shirasawa, Yamada, Aitani, Miyazaki, Matsunami, Kimura, Takahashi, Hasegawa
  et~al.}}]{HiraharaPRL109}
\bibinfo{author}{\bibfnamefont{T.}~\bibnamefont{Hirahara}},
  \bibinfo{author}{\bibfnamefont{N.}~\bibnamefont{Fukui}},
  \bibinfo{author}{\bibfnamefont{T.}~\bibnamefont{Shirasawa}},
  \bibinfo{author}{\bibfnamefont{M.}~\bibnamefont{Yamada}},
  \bibinfo{author}{\bibfnamefont{M.}~\bibnamefont{Aitani}},
  \bibinfo{author}{\bibfnamefont{H.}~\bibnamefont{Miyazaki}},
  \bibinfo{author}{\bibfnamefont{M.}~\bibnamefont{Matsunami}},
  \bibinfo{author}{\bibfnamefont{S.}~\bibnamefont{Kimura}},
  \bibinfo{author}{\bibfnamefont{T.}~\bibnamefont{Takahashi}},
  \bibinfo{author}{\bibfnamefont{S.}~\bibnamefont{Hasegawa}},
  \bibnamefont{et~al.}, \bibinfo{journal}{Phys. Rev. Lett.}
  \textbf{\bibinfo{volume}{109}}, \bibinfo{pages}{227401}
  (\bibinfo{year}{2012}),
  \urlprefix\url{http://link.aps.org/doi/10.1103/PhysRevLett.109.227401}.

\bibitem[{\citenamefont{Liu and Allen}(1995)}]{LiuAllenPRB}
\bibinfo{author}{\bibfnamefont{Y.}~\bibnamefont{Liu}} \bibnamefont{and}
  \bibinfo{author}{\bibfnamefont{R.~E.} \bibnamefont{Allen}},
  \bibinfo{journal}{Phys. Rev. B} \textbf{\bibinfo{volume}{52}},
  \bibinfo{pages}{1566} (\bibinfo{year}{1995}),
  \urlprefix\url{http://link.aps.org/doi/10.1103/PhysRevB.52.1566}.

\bibitem[{\citenamefont{Slater and Koster}(1954)}]{SlaterKoster}
\bibinfo{author}{\bibfnamefont{J.~C.} \bibnamefont{Slater}} \bibnamefont{and}
  \bibinfo{author}{\bibfnamefont{G.~F.} \bibnamefont{Koster}},
  \bibinfo{journal}{Phys. Rev.} \textbf{\bibinfo{volume}{94}},
  \bibinfo{pages}{1498} (\bibinfo{year}{1954}),
  \urlprefix\url{http://link.aps.org/doi/10.1103/PhysRev.94.1498}.

\bibitem[{\citenamefont{Lewenkopf and Mucciolo}(2013)}]{Lewenkopf}
\bibinfo{author}{\bibfnamefont{C.~H.} \bibnamefont{Lewenkopf}}
  \bibnamefont{and} \bibinfo{author}{\bibfnamefont{E.~R.}
  \bibnamefont{Mucciolo}}, \bibinfo{journal}{Journal of Computational
  Electronics} \textbf{\bibinfo{volume}{12}}, \bibinfo{pages}{203}
  (\bibinfo{year}{2013}),
  \urlprefix\url{http://dx.doi.org/10.1007/s10825-013-0458-7}.

\bibitem[{\citenamefont{Lopez-Sancho et~al.}(1985)\citenamefont{Lopez-Sancho,
  Lopez-Sancho, and Rubio}}]{SanchoRubio}
\bibinfo{author}{\bibfnamefont{M.~P.} \bibnamefont{Lopez-Sancho}},
  \bibinfo{author}{\bibfnamefont{J.~M.} \bibnamefont{Lopez-Sancho}},
  \bibnamefont{and} \bibinfo{author}{\bibfnamefont{J.}~\bibnamefont{Rubio}},
  \bibinfo{journal}{I. Phys. F: Met. Phys.} \textbf{\bibinfo{volume}{15}},
  \bibinfo{pages}{851} (\bibinfo{year}{1985}),
  \urlprefix\url{http://dx.doi.org/10.1088/0305-4608/15/4/009}.

\bibitem[{\citenamefont{Fu and Kane}(2007)}]{FuPRB}
\bibinfo{author}{\bibfnamefont{L.}~\bibnamefont{Fu}} \bibnamefont{and}
  \bibinfo{author}{\bibfnamefont{C.~L.} \bibnamefont{Kane}},
  \bibinfo{journal}{Phys. Rev. B} \textbf{\bibinfo{volume}{76}},
  \bibinfo{pages}{045302} (\bibinfo{year}{2007}),
  \urlprefix\url{http://link.aps.org/doi/10.1103/PhysRevB.76.045302}.

\bibitem[{\citenamefont{Bieniek et~al.}(2016)\citenamefont{Bieniek, Wo\'zniak,
  and Potasz}}]{acta2016}
\bibinfo{author}{\bibfnamefont{M.}~\bibnamefont{Bieniek}},
  \bibinfo{author}{\bibfnamefont{T.}~\bibnamefont{Wo\'zniak}},
  \bibnamefont{and} \bibinfo{author}{\bibfnamefont{P.}~\bibnamefont{Potasz}},
  \bibinfo{journal}{Acta Phys. Pol.} \textbf{\bibinfo{volume}{130}},
  \bibinfo{pages}{609} (\bibinfo{year}{2016}),
  \urlprefix\url{http://dx.doi.org/10.12693/APhysPolA.130.609}.

\bibitem[{\citenamefont{Han et~al.}(2014)\citenamefont{Han, Kawakami, Gmitra,
  and Fabian}}]{Fabian}
\bibinfo{author}{\bibfnamefont{W.}~\bibnamefont{Han}},
  \bibinfo{author}{\bibfnamefont{R.~K.} \bibnamefont{Kawakami}},
  \bibinfo{author}{\bibfnamefont{M.}~\bibnamefont{Gmitra}}, \bibnamefont{and}
  \bibinfo{author}{\bibfnamefont{J.}~\bibnamefont{Fabian}},
  \bibinfo{journal}{Nature Nano.} \textbf{\bibinfo{volume}{9}},
  \bibinfo{pages}{794} (\bibinfo{year}{2014}),
  \urlprefix\url{http://dx.doi.org/10.1038/NNANO.2014.214}.

\bibitem[{\citenamefont{Fratini et~al.}(2013)\citenamefont{Fratini,
  Gos\'albez-Mart\'{\i}nez, Merodio~C\'amara, and
  Fern\'andez-Rossier}}]{Fratini}
\bibinfo{author}{\bibfnamefont{S.}~\bibnamefont{Fratini}},
  \bibinfo{author}{\bibfnamefont{D.}~\bibnamefont{Gos\'albez-Mart\'{\i}nez}},
  \bibinfo{author}{\bibfnamefont{P.}~\bibnamefont{Merodio~C\'amara}},
  \bibnamefont{and}
  \bibinfo{author}{\bibfnamefont{J.}~\bibnamefont{Fern\'andez-Rossier}},
  \bibinfo{journal}{Phys. Rev. B} \textbf{\bibinfo{volume}{88}},
  \bibinfo{pages}{115426} (\bibinfo{year}{2013}),
  \urlprefix\url{http://link.aps.org/doi/10.1103/PhysRevB.88.115426}.

\bibitem[{\citenamefont{Huertas-Hernando
  et~al.}(2006)\citenamefont{Huertas-Hernando, Guinea, and Brataas}}]{Hernando}
\bibinfo{author}{\bibfnamefont{D.}~\bibnamefont{Huertas-Hernando}},
  \bibinfo{author}{\bibfnamefont{F.}~\bibnamefont{Guinea}}, \bibnamefont{and}
  \bibinfo{author}{\bibfnamefont{A.}~\bibnamefont{Brataas}},
  \bibinfo{journal}{Phys. Rev. B} \textbf{\bibinfo{volume}{74}},
  \bibinfo{pages}{155426} (\bibinfo{year}{2006}),
  \urlprefix\url{http://link.aps.org/doi/10.1103/PhysRevB.74.155426}.

\bibitem[{\citenamefont{Harrison}(1980)}]{Harrison}
\bibinfo{author}{\bibfnamefont{W.~A.} \bibnamefont{Harrison}},
  \emph{\bibinfo{title}{Electronic Structure and the Properties of Solids}}
  (\bibinfo{publisher}{Dover Publications, INC., New York},
  \bibinfo{year}{1980}), \bibinfo{edition}{1st} ed.

\end{thebibliography}

\end{document}